\documentclass[11pt, a4paper]{article}

\usepackage[utf8]{inputenc}
\usepackage[T1]{fontenc}
\usepackage[margin=1in]{geometry} 
\usepackage{graphicx}             
\usepackage{xcolor}               
\usepackage{authblk}              
\usepackage{mathpazo}             
\usepackage[scaled=0.92]{helvet}  
\usepackage{amsmath, amssymb}     
\usepackage{booktabs}             
\usepackage{cite}                 
\usepackage{fancyhdr}             
\usepackage{titlesec}             
\usepackage{lineno}               
\usepackage[colorlinks=true, urlcolor=acelleraBlue, linkcolor=acelleraBlue, citecolor=acelleraBlue]{hyperref}
\usepackage{subcaption}
\usepackage{hyperref}
\usepackage{xspace}
\definecolor{acelleraBlue}{RGB}{35, 87, 151}

\titleformat{\section}
  {\sffamily\Large\bfseries\color{acelleraBlue}}
  {\thesection}{1em}{}
  
\titleformat{\subsection}
  {\sffamily\large\bfseries\color{acelleraBlue}}
  {\thesubsection}{1em}{}

\newcommand{\pKa}{p\textit{K}\textsubscript{a}\xspace}   

\title{\sffamily\bfseries\boldmath \color{acelleraBlue} Acep$K_{\rm a}$: Thermodynamics-Informed p$K_{\rm a}$ Prediction and Protonation-State Generation in PlayMolecule AI}

\author[1]{Francesco Pesce}
\author[1]{Stephen E. Farr}
\author[1,2,3,4]{Gianni De Fabritiis}

\affil[1]{Acellera Labs, C. Dr. Trueta 183, 08005 Barcelona, Spain}
\affil[2]{Acellera, 38350 Fremont Blvd 203 Fremont CA, 94536 USA}
\affil[3]{Institució Catalana de Recerca i Estudis Avançats (ICREA), Passeig Lluis Companys 23, 08010 Barcelona, Spain}
\affil[4]{Computational Science Laboratory, Barcelona Biomedical Research Park (PRBB), Universitat Pompeu Fabra, C. Dr. Aiguader 88, 08003 Barcelona, Spain}

\date{\vspace{-5ex}} 

\begin{document}

\maketitle

\begin{abstract}
\noindent
The acid dissociation constants (\pKa) and the protonation states that they determine
govern solubility, permeability, and protein--ligand binding, making their
accurate prediction essential in drug discovery. We present Ace\pKa, an
application in the PlayMolecule AI platform that implements the Uni-\pKa
framework, which couples statistical mechanics with representation learning.
Rather than treating \pKa as a scalar regression target, Ace\pKa models the
complete protonation ensemble, enforcing thermodynamic consistency across
coupled ionization sites. The application is built on an independently
retrained Uni-Mol backbone that matches state-of-the-art accuracy on standard
public benchmarks. We
further describe three engineering contributions: AceConfgen, a
GPU-accelerated conformer generator approximately 7 times faster than
other GPU implementations and more than an order of magnitude faster than
multithreaded RDKit; a streamlined inference engine that protonates molecules
directly; and a 3D-aware mode that applies predicted protonation states to
bound ligand poses. Ace\pKa supports library-scale prediction and provides a
validated, ready-to-use implementation of this methodology, available at \url{https://open.playmolecule.org}.
\end{abstract}

\section{Introduction}
The acid dissociation constant (p$K_{\rm a}$) is a fundamental physicochemical parameter that describes the equilibrium between protonated and deprotonated forms of a molecule. In drug discovery, p$K_{\rm a}$ is paramount because it governs a compound's ionization state under physiological conditions. This, in turn, dictates critical pharmacokinetic properties including aqueous solubility, membrane permeability (lipophilicity), and distribution \cite{manallack2013significance}. Furthermore, the ionization state determines the specific electrostatic interactions a ligand can form with a protein target, directly impacting binding affinity and efficacy.

Despite its importance, the prediction of p$K_{\rm a}$ values remains non-trivial. The difficulty arises from the complexity of molecular electronic structures. Many drug-like molecules are polyprotic, containing multiple ionizable sites that influence each other through inductive and resonance effects. This leads to coupled equilibria where the ionization of one group shifts the p$K_{\rm a}$ of neighbors.

Traditional computational approaches often struggle to resolve this complexity: template-based methods and early QSAR models rely heavily on empirical corrections and local atomic descriptors, which limit their generalizability across diverse chemical spaces~\cite{mansouri2019open}. While recent deep learning advancements, such as graph neural networks (e.g., MolGpKa, pKasolver), have improved predictive capabilities, they typically frame $pK_a$ prediction as a site-specific regression task~\cite{pan2021molgpka, mayr2022improving}. This simplification often neglects the global protonation network, leading to thermodynamic inconsistencies where predicted constants violate fundamental thermodynamic cycles~\cite{gunner2020standard}. Conversely, quantum mechanical (QM) methods, though physically rigorous, entail prohibitive computational costs necessitated by exhaustive conformational sampling and solvent modeling~\cite{bochevarov2016multiconformation, yu2018weighted}. To bridge this gap, Uni-p$K_{\rm a}$ introduced a unified framework that integrates thermodynamic principles with the Uni-Mol 3D molecular representation learner~\cite{zhou2023uni, luo2024bridging}. By reconstructing the complete ``protonation ensemble'' via a protonation-state enumerator and predicting free energies rather than $K_{\rm a}$ values directly, Uni-p$K_{\rm a}$ inherently preserves thermodynamic consistency. This approach not only achieves state-of-the-art accuracy on public benchmarks but also provides a robust interpretation of experimental macro-$pK_a$ data that rivals QM-based methods at a fraction of the computational cost.

Drawing from this recently proposed framework, we introduce Acep$K_{\rm a}$. This application builds upon the rigorous theoretical foundation of Uni-p$K_{\rm a}$ while introducing practical enhancements in computational efficiency and usability. Acep$K_{\rm a}$ is integrated into the PlayMolecule AI platform \cite{navarro2025speak}, which removes any accessibility barrier and makes it directly exploitable in drug discovery/design workflows to serve the practical needs of medicinal chemists.

While the Uni-p$K_{\rm a}$ methodology is available in an open-source repository, the trained models underlying its published benchmarks are not distributed, and its \href{https://www.bohrium.com/notebooks/38543442597}{public demo} is not adequate for library-scale processing. Obtaining the published accuracy therefore requires retraining the model, which demands curated data, GPU resources, and expertise. Acep$K_{\rm a}$ removes these barriers. To our knowledge, Acep$K_{\rm a}$ is the first freely accessible, library-scale implementation of the Uni-$K_{\rm a}$ framework.

\section{Methods}
To make the scope of this work explicit, we distinguish the elements inherited from the underlying frameworks from those introduced here. Acep$K_{\rm a}$ inherits the protonation-ensemble formalism (Eqs.~\ref{eq:ka_def}--\ref{eq:boltz_weights}), the rule-based microstate enumerator, and the Uni-Mol representation backbone from Uni-p$K_{\rm a}$ and Uni-Mol~\cite{luo2024bridging, zhou2023uni}. The contributions of the present work are an independently retrained and publicly usable model, the AceConfgen GPU-native conformer engine, a streamlined library-scale inference pipeline, a 3D bound-pose modality, and a deployment within PlayMolecule AI.

\subsection{Microstate Population and Free Energy}
Acep$K_{\rm a}$ builds on the same foundation as Uni-p$K_{\rm a}$, establishing a rigorous relationship between protonation states and free energies. In the Brønsted-Lowry acid-base theory, a molecule exists as a protonation ensemble, \textit{i.e.} a distribution of protonated forms.

The micro-p$K_{\rm a}$ characterizes the equilibrium between two specific protonation states (microstates), while the macro-p$K_{\rm a}$ represents the equilibrium between the ensemble of protonated species (macrostate $\rm HA$) and the ensemble of deprotonated species (macrostate $\rm A^-$).

The macroscopic acid dissociation constant ($K_{\rm a}$) for an acid-base equilibrium is derived from the ratio of the partition functions of the base and acid macrostates. Given that the population of any microstate in the ensemble follows the Boltzmann distribution based on its standard/dimensionless free energy ($G$), $K_{\rm a}$ can be expressed as:

\begin{equation}
    K_a = \frac{[{\rm H^+}]\sum_{i \in {\rm A^-}} [{\rm A^-_i}]}{\sum_{j \in \rm HA} [{\rm HA}]_j} = \frac{\sum_{i \in {\rm A^-}} e^{-G_i/RT}}{\sum_{j \in \rm{HA}} e^{-G_j/RT}}
    \label{eq:ka_def}
\end{equation}
where $G_i$ and $G_j$ are the dimensionless Gibbs free energies of the deprotonated ($i \in A^-$) and protonated ($j \in HA$) microstates, respectively, $R$ is the universal gas constant and $T$ is the temperature.

Since the relationship between the population of the microstates and their free energies is governed by the Boltzmann distribution, at a given pH the Boltzmann weight of a microstate can be calculated as:

\begin{equation}
    w_i(pH)=\frac{\exp(-G_i-q_i\ln(10)\cdot \rm pH)}{\sum_{j=1}^N\exp(-G_j-q_j\ln(10)\cdot \rm pH)}
    \label{eq:boltz_weights}
\end{equation}
where $q_i$ is the net charge of the microstate $i$. The denominator of Eq. \ref{eq:boltz_weights} represents the partition function of the protonation ensemble at the query pH, summing over all $N$ possible microstates

By predicting $G$ for every microstate, Acep$K_{\rm a}$ can analytically compute the macro-p$K_{\rm a}$s, the population of microspecies at any given pH, guaranteeing that the results are thermodynamically consistent.

\subsection{Architecture}
The core predictive engine of Acep$K_{\rm a}$ is based on the Uni-p$K_{\rm a}$ architecture. This framework represents a paradigm shift from direct property prediction to free energy modeling. The architecture consists of three main components:
\begin{enumerate}
    \item \textbf{Microstate Enumerator:} A rule-based module that generates a comprehensive protonation ensemble for an input molecule in any generic microstate. This module exploits a template containing SMARTS patterns to identify ionizable sites and combinatorially generate all valid microstates with net charge bounded between -2 and +2. Different templates are available, tailored to handle ionizable groups common in drug-like compounds, as well as more general cases.
    \item \textbf{Uni-Mol Backbone:} A 3D molecular representation learning framework that utilizes a Transformer-based architecture invariant to SE(3) transformations (rotation and translation) \cite{zhou2023uni}. It takes as input the coordinates and atom types for a molecule and predicts its standard Gibbs free energy. It employs self-attention layers to capture crucial non-local atomic interactions and spatial electronic features.
    \item \textbf{FE2p$K_{\rm a}$ Module:} Free energies predicted by Uni-Mol for each microstate of a protonation ensemble are fed to this module that employs Eq. \ref{eq:ka_def} and Eq. \ref{eq:boltz_weights} to return macro-p$K_{\rm a}$ values and relative pH-dependent microstate populations.
\end{enumerate}

\section{Results}

\subsection{Re-training and Model Evaluation}
Acep$K_{\rm a}$ has been re-trained following the Uni-p$K_{\rm a}$ protocol \cite{luo2024bridging}, to reproduce its accuracy and generalizability across chemical space. The Uni-Mol model \cite{zhou2023uni} was pre-trained on a weakly supervised and three self-supervised tasks. The supervised task consists of training to predict p$K_{\rm a}$ values from the ChEMBL database, leveraging approximately 1 million molecules with empirical p$K_{\rm a}$ values, which expanded to over 3 million unique protonation states after microstate enumeration. Using self-supervised tasks, masked atom prediction, 3D coordinate recovery, and masked charge prediction, the model acquires a general ”understanding” of chemical features and geometries.

\begin{figure}[h]
    \centering
    \includegraphics[width=\textwidth]{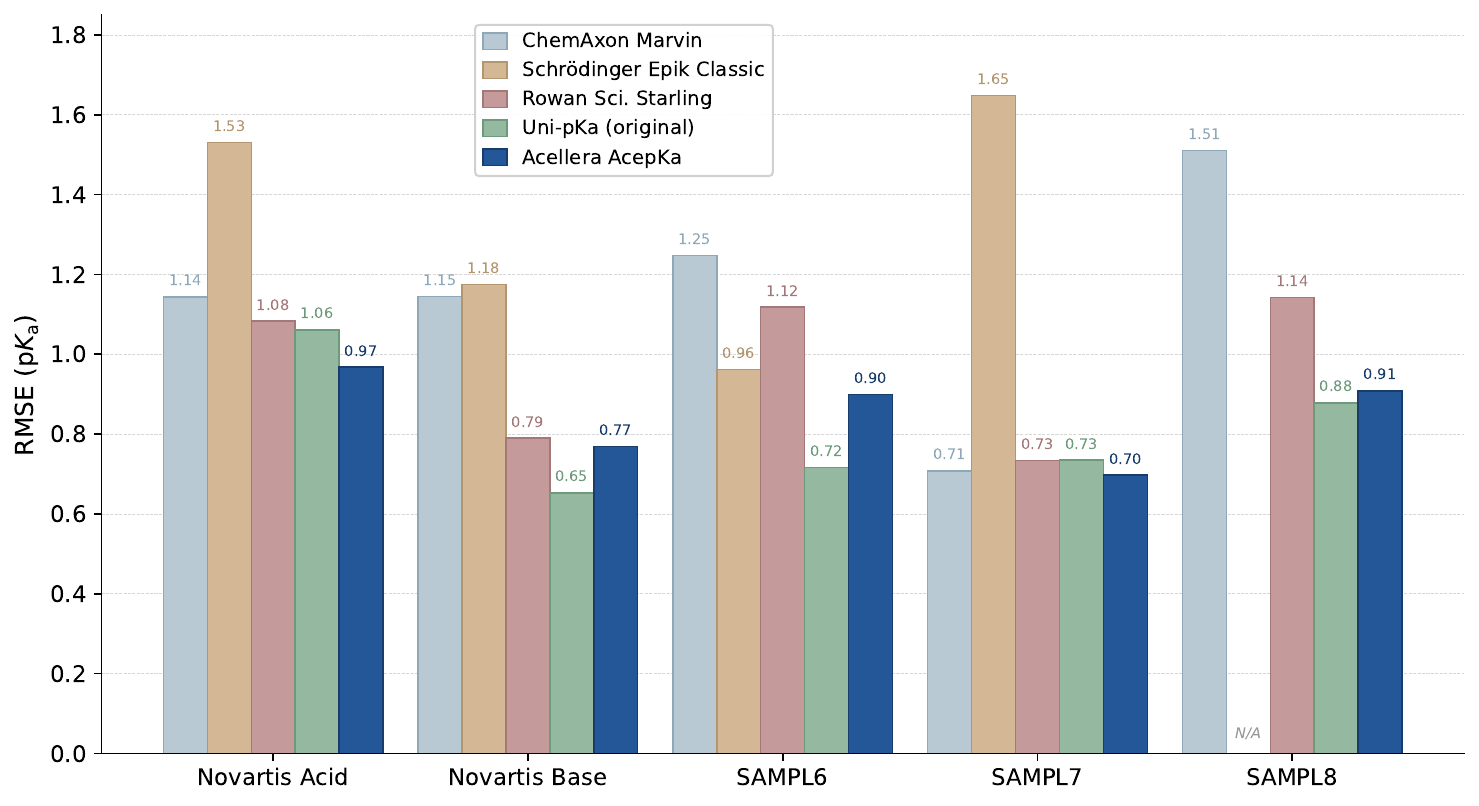} 
    \caption{Benchmark of macro-p$K_{\rm a}$ prediction accuracy (RMSE, in p$K_{\rm a}$ units; lower is better) on public datasets. Acep$K_{\rm a}$ values were computed in this work; all other values are taken from~\cite{luo2024bridging} and~\cite{wagen2025physics}.}
    \label{fig:barplot}
\end{figure}

Following pre-training, the model is fine-tuned on high-quality public p$K_{\rm a}$ datasets from DataWarrior and selected curated entries from the i-BonD database. Fine-tuning produced a 5-fold ensemble model. On standard public benchmarks, Acep$K_{\rm a}$ achieves accuracy on par with the state of the art (Fig.~\ref{fig:barplot}), performing comparably to the original Uni-p$K_{\rm a}$ results and to leading commercial tools. Because Acep$K_{\rm a}$ and Uni-p$K_{\rm a}$ share the same underlying methodology, close agreement is the expected outcome, and it indicates that our independent retraining faithfully reproduces the published approach in a directly usable form. The remaining differences (Acep$K_{\rm a}$ is more accurate on some datasets and marginally less on others) are within the statistical uncertainty of these relatively small benchmark sets and are attributable to differences in fine-tuning data composition and implementation rather than to a systematic advantage of either model. Notably, Acep$K_{\rm a}$ makes this ensemble model directly usable, whereas the ensemble underlying the published Uni-p$K_{\rm a}$ results is not distributed.

\subsection{App Workflow and Capabilities}

\begin{figure}[h]
    \centering
    \includegraphics[width=\textwidth]{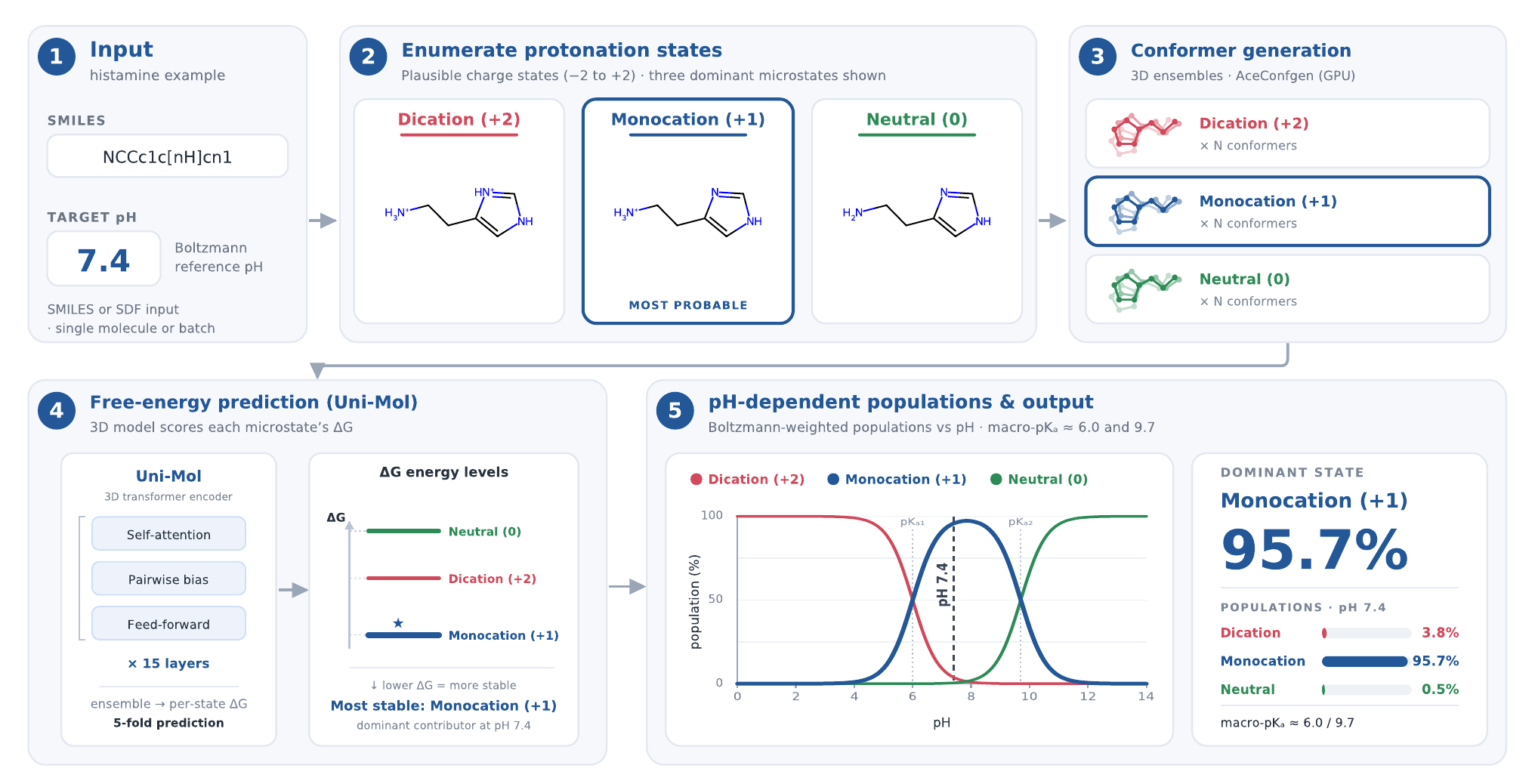}
    \caption{Acep$K_{\rm a}$ app workflow. (1) The app takes as input one or multiple SMILES strings (or 3D conformers) and a pH value. (2) It then generates the protonation ensemble for each of the provided molecules. (3) A conformational ensemble for each microstate is generated using AceConfgen. (4) The 3D conformations are fed to the Uni-Mol model, which predicts their free energies. (5) The free energies are used in Eq.~\ref{eq:boltz_weights} to derive the relative pH-dependent population of each microstate.}
    \label{fig:workflow}
\end{figure}

The Acep$K_{\rm a}$ app structure (Fig. \ref{fig:workflow}) follows the Uni-p$K_{\rm a}$ workflow, introducing several functional enhancements over the original code to support high-throughput and structure-based drug design workflows:

\begin{itemize}
    \item \textbf{Streamlined, library-scale inference:} The inference pipeline chains microstate enumeration, batched GPU conformer generation (via AceConfgen; see below), and batched Uni-Mol free-energy evaluation into a single end-to-end process.
    \item \textbf{Accelerated Conformer Generation:} Accurate p$K_{\rm a}$ prediction requires reasonable ensembles of 3D conformers for the microstates. To achieve this at scale, we utilize an in-house GPU-accelerated engine (AceConfgen) which, running on a consumer GPU, is roughly 7$\times$ faster than the GPU competitor \texttt{nvmolkit} (\url{https://github.com/NVIDIA-Digital-Bio/nvMolKit}) and more than an order of magnitude faster than CPU-based RDKit. This code significantly reduces the computational overhead of generating the dataset that serves as input to the Uni-Mol model.
    \item \textbf{3D Modality for Bound Poses:} Acep$K_{\rm a}$ supports a ``3D modality'' beyond standard SMILES input/output. It can accept a small molecule in a specific bound pose (e.g., from a crystal structure or docking pose) and return that same pose with the predicted protonation and hydrogen placement applied in place. We emphasize that the protonation state is predicted from the ligand alone: the protein environment is not used as an input, and pocket-induced p$K_{\rm a}$ shifts are not modeled. The value of this modality is coordinate preservation.
\end{itemize}

\begin{figure}[h!]
    \centering
    \includegraphics[width=3.25in]{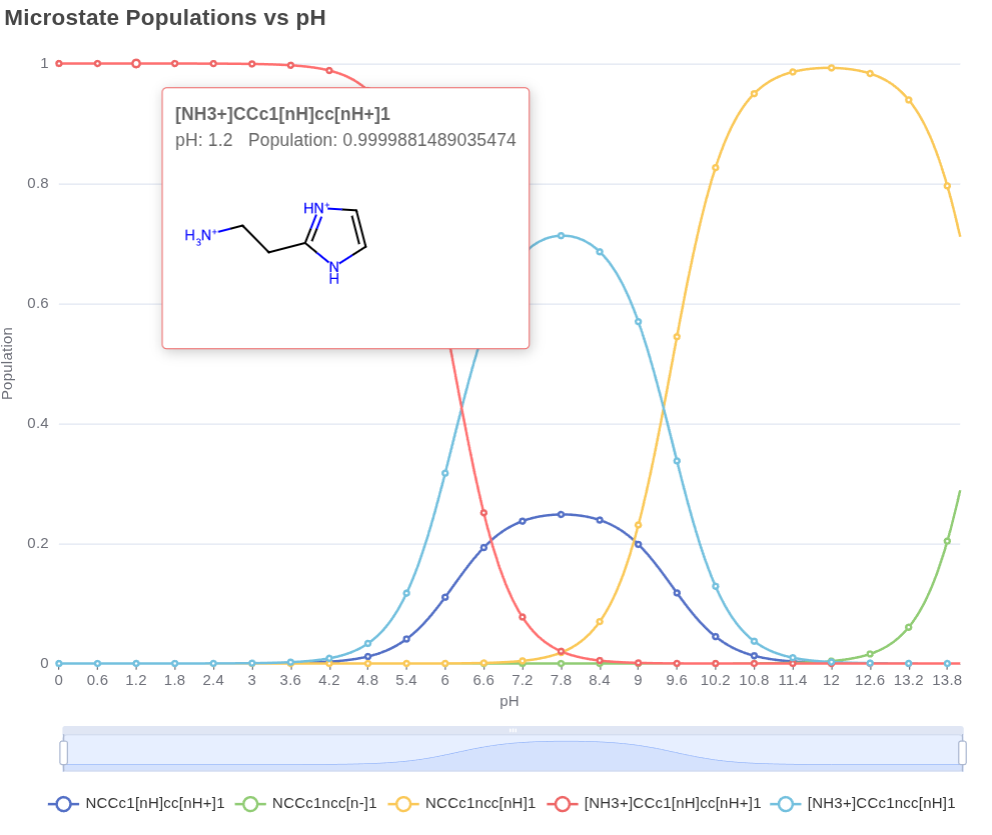} 
    \caption{Output example from Acep$K_{\rm a}$ in single-molecule mode using histamine as query molecule. The figure shows the relative pH-dependent population of each enumerated microstate.}
    \label{fig:single_mol_mode}
\end{figure}

The application can run in two modes:
\begin{itemize}
    \item \textbf{Single-molecule mode}: if a single molecule is passed to the app for prediction, the app will return (i) macro-p$Ka$ values, (ii) relative pH-dependent population of each microstate (Fig. \ref{fig:single_mol_mode}), (iii) molecule in its highest populated state at the queried target pH.
    \item \textbf{Library mode}: if multiple molecules are provided, the app will return the highest populated protonated state for each molecule at the queried target pH.
\end{itemize}
Both modes can be executed on SMILEs and 3D conformations (from SDF files).

\subsection{Accelerated conformer generation - AceConfgen}

\begin{figure}[htbp]
     \centering
     \includegraphics[width=\textwidth]{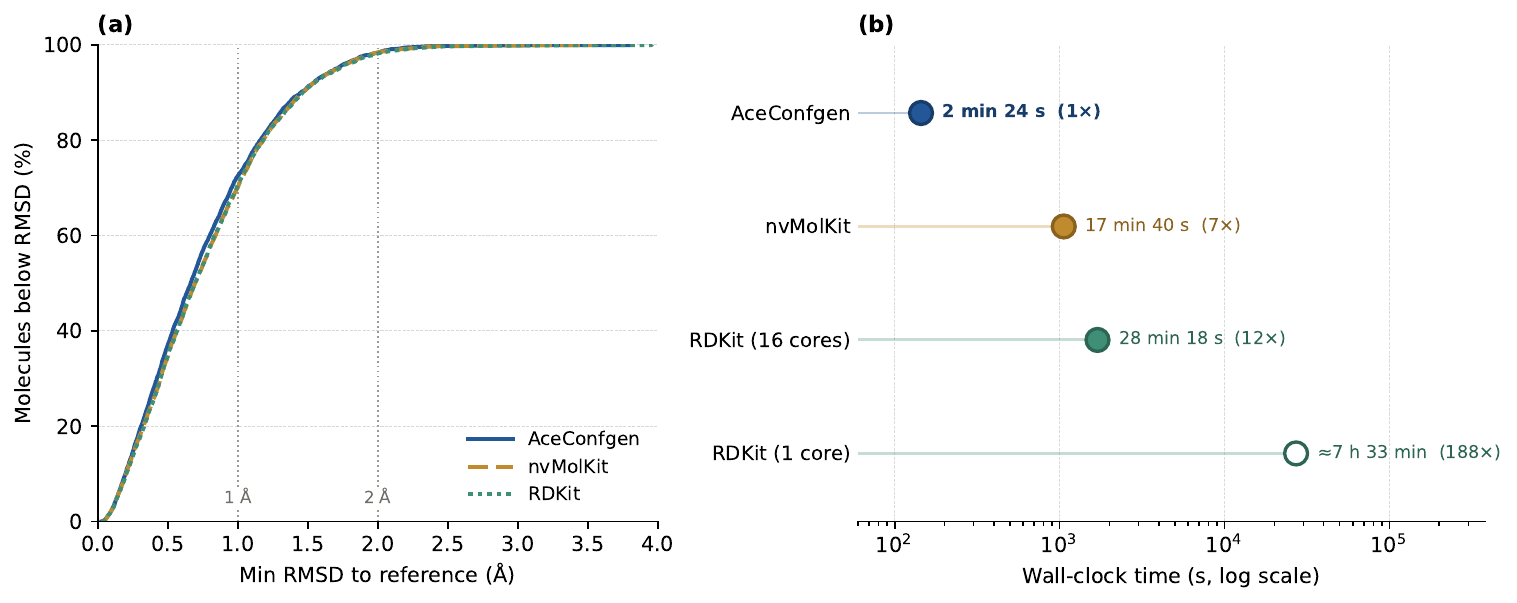}
     \caption{Conformer-generation benchmark on the Platinum 2017 dataset (50 conformers per molecule for 4{,}548 molecules; GPU methods on a single NVIDIA RTX 4090). \textbf{(a)} Empirical cumulative distribution of the per-molecule minimum RMSD to the reference bioactive pose; AceConfgen, nvMolKit and RDKit reach essentially identical accuracy. \textbf{(b)} Wall-clock time to generate all 227{,}400 conformers (log scale); the RDKit single-core point is extrapolated as $16\times$ the 16-core run. AceConfgen is $\sim$7$\times$ faster than nvMolKit and $\sim$12$\times$ faster than 16-core RDKit. Single-thread runtime of RDKit extrapolated for reference as 16 times the runtime of 16-core RDkit.}
     \label{fig:conf_benchmark}
\end{figure}

As part of the Acep$K_{\rm a}$ suite, we developed AceConfgen, a GPU-native implementation for conformer generation. AceConfgen is an integrated engineering component of Acep$K_{\rm a}$ rather than a new algorithm: it reproduces RDKit's standard distance-geometry and MMFF94 minimization procedures using bespoke fused kernels for all force, energy, and minimization routines. Because ensemble-based p$K_{\rm a}$ prediction requires 3D conformers for every enumerated microstate of every molecule, conformer generation can be the dominant cost of the pipeline at library scale, and accelerating it is what makes the overall method practical on commodity hardware. By utilizing FP32 precision, AceConfgen achieves high throughput on consumer-grade hardware, such as the NVIDIA RTX 4090. In contrast, the alternative software, nvMolKit, utilizes FP64 for specific operations, which significantly limits its performance on consumer GPUs. To validate accuracy, we evaluated AceConfgen against the Platinum 2017 benchmark dataset\cite{friedrich2017}, which consists of 4,548 molecules with reference conformers derived from protein-ligand structures. The benchmark objective is to generate $N$ conformers per molecule and calculate the Root Mean Square Deviation (RMSD) relative to the reference; a lower RMSD distribution indicates a higher likelihood of capturing the bioactive pose, increasing utility in drug discovery. We generated 50 conformers per molecule for all 4{,}548 molecules (227{,}400 conformers in total) with AceConfgen, nvMolKit (version 0.4.0), and RDKit, running each tool with its default settings and requesting the same number of conformers to ensure a fair comparison. The two GPU tools (AceConfgen and nvMolKit) were run on a single NVIDIA RTX 4090, whereas RDKit was run on CPU using 16 threads, with the corresponding single-thread runtime extrapolated for reference. AceConfgen, nvMolKit and RDKit reach essentially identical conformational accuracy (Fig.~\ref{fig:conf_benchmark}a), but their runtimes differ substantially (Fig.~\ref{fig:conf_benchmark}b): AceConfgen completed the benchmark in 2~min 24~s, compared with 17~min 40~s for nvMolKit and 28~min 18~s for 16-core RDKit. This corresponds to a $\sim$7$\times$ speed-up over nvMolKit and a $\sim$12$\times$ speed-up over multi-threaded RDKit; relative to the widely used single-threaded RDKit workflow, extrapolated to $\sim$7.5~h, the speed-up approaches two orders of magnitude. Furthermore, AceConfgen successfully processed all 4{,}548 molecules, whereas nvMolKit and RDKit each failed on two.

\subsection{Implementation in PlayMolecule AI}
\begin{figure}[h!]
    \centering
    \begin{subfigure}[b]{1.\textwidth}
        \centering
        \caption{}
        \includegraphics[width=\textwidth]{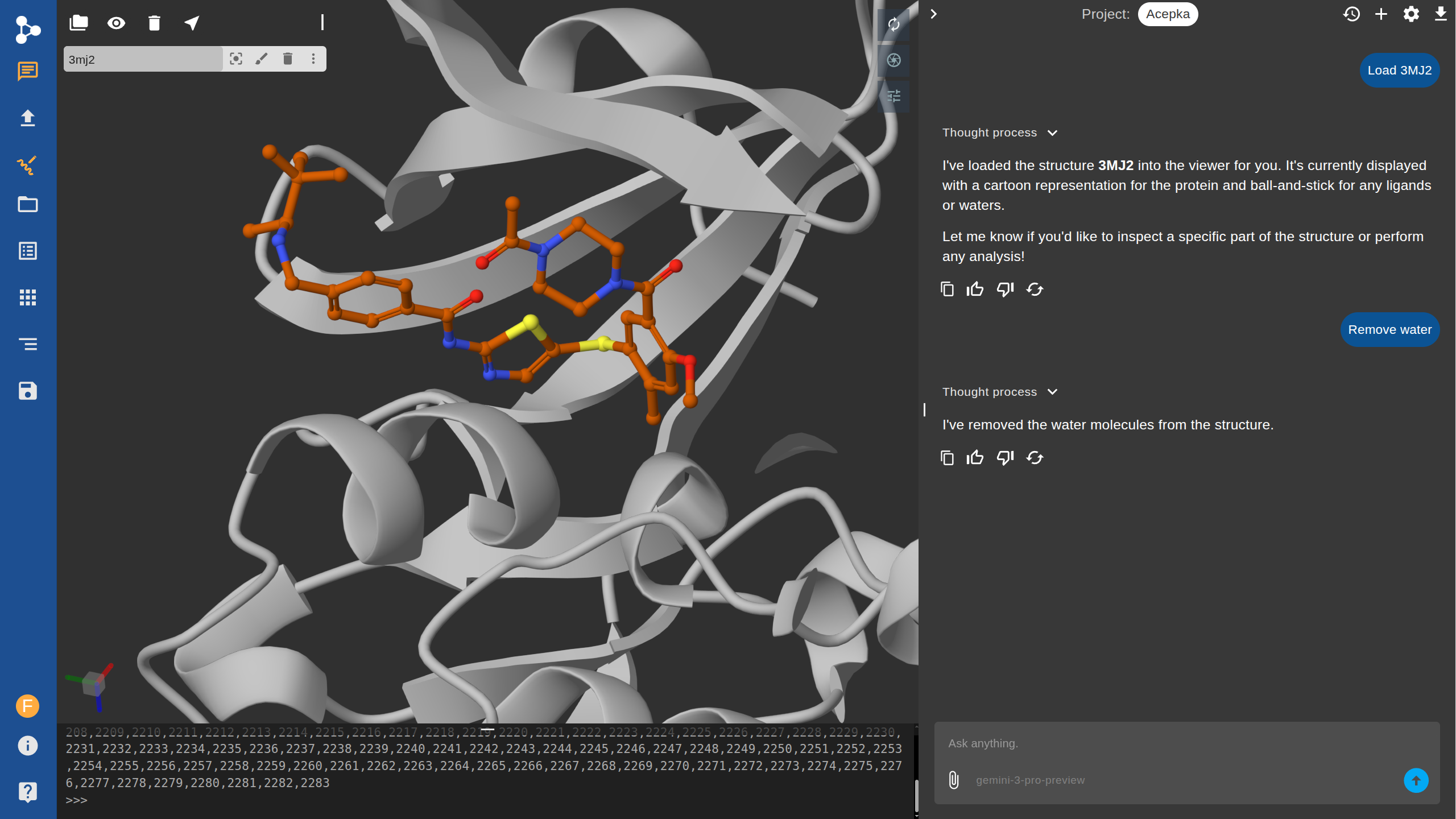} 
        \label{fig:sub1}
    \end{subfigure}
    \hfill 
    \begin{subfigure}[b]{1.\textwidth}
        \centering
        \caption{}
        \includegraphics[width=\textwidth]{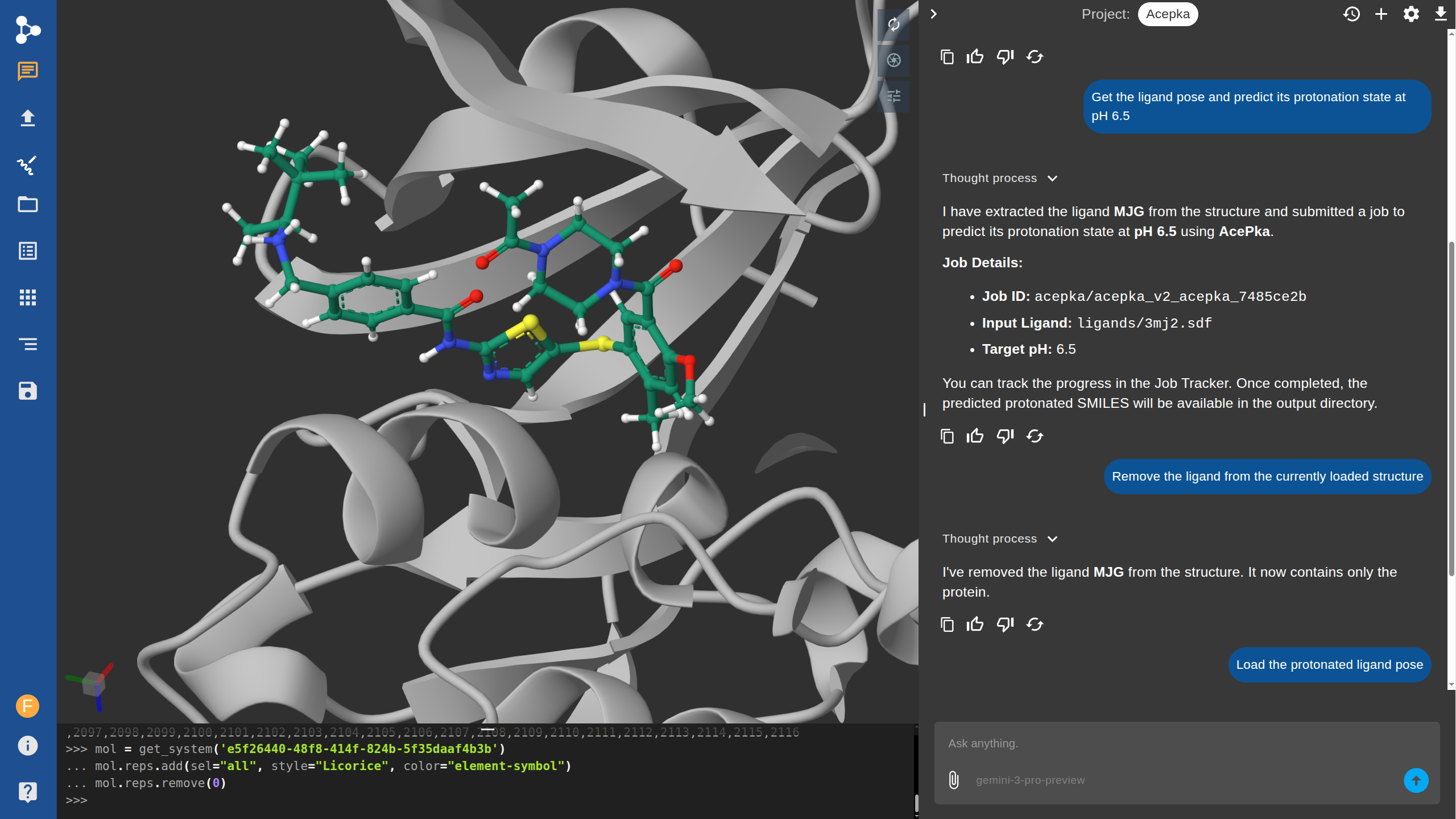}
        \label{fig:sub2}
    \end{subfigure}
    
    \caption{PlayMolecule AI GUI and Acep$K_{\rm a}$ usage orchestrated by the LLM agent, showing the prediction of the protonated state of a ligand sitting in a protein pocket. (a) The agent is asked to fetch a structure with PDB ID 3MJ2. (b) The agent is asked to predict the protonation state of the ligand present in the loaded structure at a target pH, and to visualize the results.}
    \label{fig:3}
\end{figure}

Acep$K_{\rm a}$ is deployed as a fully integrated application within \href{PlayMolecule AI}{https://open.playmolecule.org/}~\cite{navarro2025speak}. This platform combines a web-based molecular viewer with an intelligent LLM agent that acts as a ``co-scientist'' and a full suite of tools for molecular modeling and drug discovery (Fig. \ref{fig:3}).

Within this ecosystem, Acep$K_{\rm a}$ serves as a seamless utility. Users can interact with the app via natural language through the chat interface; details of the platform and its agent are given in ~\cite{navarro2025speak}. The integration allows for:
\begin{itemize}
    \item \textbf{Agent Orchestration:} The PlayMolecule AI agent can autonomously call Acep$K_{\rm a}$ as part of complex workflows, automatically preparing ligands for docking and other downstream tasks, or correcting protonation states of ligands in a PDB file before molecular dynamics simulation.
    \item \textbf{Direct Visualization:} Predicted protonation states can be immediately loaded into the viewer, allowing users to inspect hydrogen bonding networks and electrostatic complementarity with the protein target.
    \item \textbf{Decision Support:} The agent can interpret Acep$K_{\rm a}$'s output to provide recommendations. For example, if Acep$K_{\rm a}$ predicts a p$K_{\rm a}$ close to physiological pH (e.g., 7.2), the Agent can flag this to the user, suggesting that both protonated and deprotonated species might be relevant for binding.
\end{itemize}
The LLM agent is an optional convenience layer: Acep$K_{\rm a}$ can be run directly through the PlayMolecule interface without any agent interaction, and every prediction is fully accessible through this direct route.

\section{Conclusion}
Acep$K_{\rm a}$ packages the thermodynamically rigorous Uni-p$K_{\rm a}$ methodology into a freely accessible, library-scale tool. By coupling the accuracy of the Uni-Mol model with a GPU-native conformer engine and the user-centric design of PlayMolecule AI, we provide a validated, ready-to-use implementation that is both scientifically robust and practically accessible. The ability to predict ensemble properties and to apply protonation states to 3D bound poses positions Acep$K_{\rm a}$ as a versatile utility for modern computational drug discovery campaigns.

\newpage

\newpage
\section{Author information}

\subsection{Author contribution}
F.P. and G.D.F. designed the project. F.P, S.E.F. and G.D.F wrote the manuscript. F.P. trained the Acep$K_{\rm a}$ model, wrote and implemented the app. S.E.F. developed AceConfgen.

\subsection{Notes}
The authors have an economic interest in Acellera.

\section{Data and Software Availability}
Acep$K_{\rm a}$ is freely available as a web application within PlayMolecule AI at \url{https://open.playmolecule.org/}. As a browser-based service, it is operating-system agnostic and requires no local installation; use requires a free user account. Acep$K_{\rm a}$ supports both single-molecule and library (multi-molecule) prediction and accepts either SMILES strings or 3D structures (SDF files); predictions can be run directly through the web interface, without the platform's optional LLM agent. A programmatic API is not currently offered. The training data are available from~\cite{luo2024bridging} at \newline\url{https://www.aissquare.com/datasets/detail?pageType=datasets&name=Uni-pKa-Dataset}.

\clearpage
\bibliographystyle{unsrt} 
\bibliography{references}

\end{document}